\documentclass[twocolumn,prl,showpacs]{revtex4}
\usepackage{psfrag}
\usepackage{graphicx}
\usepackage{amsmath}
\usepackage{amssymb}

\begin{document}
\title{Conduction Mechanism in a Molecular Hydrogen Contact}
\author{K. S. Thygesen}
\affiliation{Center for Atomic-scale Materials Physics, \\
  Department of Physics, Technical University of Denmark, DK - 2800
  Kgs. Lyngby, Denmark}
\author{K. W. Jacobsen}
\affiliation{Center for Atomic-scale Materials Physics, \\
  Department of Physics, Technical University of Denmark, DK - 2800
  Kgs. Lyngby, Denmark}
\date{\today}

\begin{abstract}
  We present first principles calculations for the conductance of a
  hydrogen molecule bridging a pair of Pt electrodes. The transmission
  function has a wide plateau with $T\approx 1$ which extends across
  the Fermi level and indicates the existence of a single, robust
  conductance channel with nearly perfect transmission. Through a
  detailed Wannier function analysis we show that the $\text{H}_2$
  bonding state is not involved in the transport and that the plateau
  forms due to strong hybridization between the $\text{H}_2$
  anti-bonding state and states on the adjacent Pt atoms. The Wannier
  functions furthermore allow us to derive a resonant-level model for
  the system with all parameters determined from the fully
  self-consistent Kohn-Sham Hamiltonian.
\end{abstract}

 \maketitle The study of electron
transport through single molecules has evolved during the last decade
as new experimental techniques have made it possible to produce
atomic-scale contacts with a few or even a single molecule suspended
between macroscopic
electrodes~\cite{joachim95,reed97,reichert_weber02,tao03}.  At the
same time theoretical efforts have been made to describe and
understand the experiments from first
principles~\cite{taylor02,damle01,diventra02}. The connection between
experiment and theory, however, has been complicated by the crucial
but in practice uncontrollable atomistic details of the contact region
between the molecule and the leads.  While the majority of previously
investigated molecules have shown a conductance much lower than the
quantum unit, $G_0=2e^2/h$, Smit.~\emph{et al.} recently measured a
conductance close to $1G_0$ for a hydrogen molecule bridging a pair of
Pt electrodes~\cite{smit02}. The result immediately raises the
question: how can a hydrogen molecule which has a closed shell
configuration and a large energy gap be conducting? Despite the
simplicity of the system, there are still considerable disagreements
among the reported calculations for the conductance of the hydrogen
bridge.  Quantitatively, values of
$0.9G_0$~\cite{smit02,cuevas_heurich03} and
$(0.2-0.5)G_0$~\cite{garcia_palacios04} have been published by
different group using similar methods.  Perhaps even more importantly,
the physical explanations for the obtained results are very different.
Indeed, both the bonding~\cite{cuevas_heurich03,garcia_palacios04} as
well as the anti-bonding~\cite{smit02} state of the $\text{H}_2$
molecule have been proposed as the current-carrying state.

In this letter we present conductance calculations based on Density
Functional Theory (DFT) showing that a hydrogen molecule bridging a
pair of Pt contacts can have a conductance close to $1G_0$ and we explain the physical mechanism
behind this result. The transmission function is found to have a characteristic
plateau with $T\approx 1$ in an energy window of 4 eV around the Fermi
level, indicating the existence of a single, very robust conductance
channel with nearly perfect transmission.
By performing a Wannier function (WF) analysis we can
directly study the transmission through the $\text{H}_2$ bonding and
anti-bonding states separately.
The results clearly demonstrate that the bonding
state takes almost no part in the transport and that the plateau is a
result of a strong hybridization between the $\text{H}_2$ anti-bonding
state and a combination of $d$- and $s$-like orbitals located on
the neighboring Pt atoms. The analysis furthermore
allows us to determine characteristic model parameters from first
principles which in turn provides a very simple description of the system.

\begin{figure}[!b]
\includegraphics[width=1.0\linewidth]{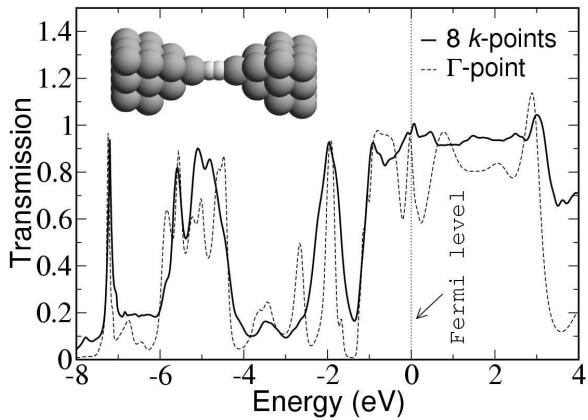}
\caption[cap.wavefct]{\label{fig1} Calculated transmission for the
  molecular hydrogen contact shown in the inset. For comparison both the $\bold
  k$-point sampled transmission and the $\Gamma$-point transmission are
  shown. The $\Gamma$-point transmission has more structure, however, the qualitative
  features of the curves are essentially the same. The wide
  plateau with $T\approx 1$ extending across the Fermi level indicates
  a single, very robust conductance channel with nearly perfect transmission.}  \end{figure}

To describe the molecular contact we use the supercell shown in the
inset of Fig.~\ref{fig1}. It contains the $\text{H}_2$ molecule
anchored between two 4-atom Pt pyramids which again are attached to
Pt(111) surfaces~\cite{calc}. We calculate the conductance of the
relaxed structures assuming that the electrons move phase coherently
through the contact and are influenced only by the self-consistent
Kohn-Sham potential.  In this case the conductance is given by
$G=G_0T(\varepsilon_F)$, where $T(\varepsilon_F)$ is the transmission
function at the Fermi level~\cite{datta_book}. The transmission
function is found using the Green's function method described in
Refs.~\cite{thygesen_bollinger03,datta02,brandbyge02}. In this
approach the system is divided into three regions: A left lead, $L$, a
right lead, $R$, and a central region, $C$. The leads are assumed to
be periodic such that all scattering takes place in $C$. In our case
$C$ coincides with the supercell of the DFT calculation and the leads
are bulk Pt(fcc) described in a supercell containing $3\times 3$ atoms
in the transverse plane to match the central region at the interfaces.
The transmission function is then given by the
formula~\cite{meirwingreen}
\begin{equation}\label{eq.condformula}
T(\varepsilon)=\text{Tr}\big [G^r_C(\varepsilon) \Gamma_L(\varepsilon) G^a_C(\varepsilon)\Gamma_R(\varepsilon)\big],
\end{equation}  
where $G_C^r$ is the retarded Green's function of the scattering region
\begin{equation}
G^r_C(\varepsilon)=[(\varepsilon+i\eta^{+})S-\Sigma_L(\varepsilon)-\Sigma_R(\varepsilon)-H_C]^{-1}.
\end{equation}
Here $H_C$ and $S$ are the Hamiltonian and overlap
matrices of the central region, $\eta^+$ is a positive infinitesimal and 
$\Sigma_{\alpha}$ is the self-energy from lead $\alpha$. The coupling strength of lead
$\alpha$ is given by
$\Gamma_{\alpha}=i(\Sigma_{\alpha}-\Sigma_{\alpha}^{\dagger})$.

We use partly occupied Wannier functions, $\{\phi_{n\alpha}\}$, (see
below) as basis functions in each of the three regions
($\alpha=L,R,C$).  Due to the limited size of the
supercell in the plane perpendicular to the transport direction the
conductance should be calculated as an integral over the Brillouin
zone in the corresponding plane. We thus form the Bloch states
$\psi_{\bold k_{\perp}n\alpha}(\bold r)=\sum_{\bold
  R_{\perp}}e^{i\bold k_{\perp}\cdot \bold
  R_{\perp}}\phi_{n\alpha}(\bold r-\bold R_{\perp})$, where $\bold
R_{\perp}$ runs over supercells in the transverse plane.  For each
$\bold k_{\perp}$ we obtain a Hamiltonian matrix $H(\bold
k_{\perp})_{n\alpha,m\beta}=\langle \bold k_{\perp} n\alpha|H|\bold
k_{\perp}m\beta\rangle$ which in turn leads to a conductance $G(\bold
k_{\perp})$ through Eq.~(\ref{eq.condformula}). The integrated
conductance can then be approximated by the finite sum $\sum_{\bold
  k_{\perp}}w(\bold k_{\perp})G(\bold k_{\perp})$, where $w(\bold
k_{\perp})$ are appropriate weight factors. 

We focus on a single, fully relaxed contact characterized by the bond
lengths $d_{\text{H-H}}=1.0$~\AA~and $d_{\text{Pt-H}}=1.76$~\AA. 
The vibrational modes of the hydrogen molecule in this configuration
are in fair agreement with new experimental results~\cite{djukic04}.
In Fig.~\ref{fig1} we show the transmission function calculated using
8 irreducible $\bold k$-points to sample the transverse BZ. The same
curve calculated within the widely used $\Gamma$-point approximation
is shown for comparison. The two curves have essentially the same
features, however, the $\Gamma$-point curve has more structure. This
is because the $\bold k$-point sampling provides the
correct smearing of the electronic structure in the leads which effectively
washes out features related to single points in the
transverse plane of the lead Brillouin zone. An interesting feature of
the transmission function is the wide plateau with $T\approx 1$
extending across the Fermi level. We shall refer to this plateau as the $1G_0$-plateau.

\begin{figure}[!b]
\includegraphics[width=0.77\linewidth,angle=270]{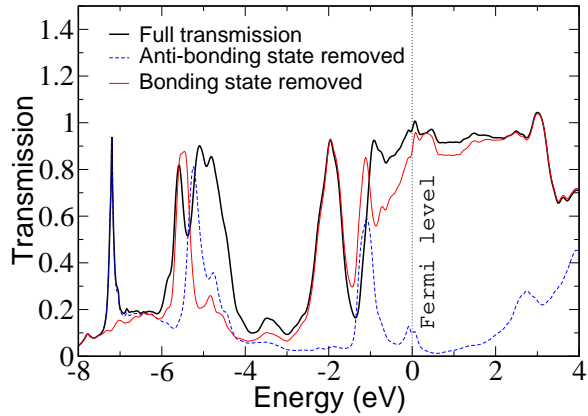}
\caption[cap.wavefct]{\label{fig2} Calculated transmission for the
  structure shown in the inset of Fig.~\ref{fig1} when all coupling to
  the bonding, respectively, the anti-bonding $\text{H}_2$ state has
  been cut. The full transmission has been repeated for
  comparison. The narrow peak around $-7$~eV is clearly due to the
  bonding state, while the peak at $-2$~eV and the wide plateau around
  the Fermi level almost
  exclusively involve the anti-bonding state.}
\end{figure}

To gain insight into the formation of the $1G_0$-plateau we perform a
Wannier function analysis. The WFs are defined as linear combinations
of the Kohn-Sham eigenstates with the expansion coefficients chosen to
make the WFs orthogonal and maximally localized. By including selected
unoccupied eigenstates in this construction we can obtain good
localization properties of the WFs also for metallic
systems~\cite{souza01, partlyoccwfs}.  We stress that the minimal WF
basis set retains the accuracy of the plane wave DFT calculation since
the WFs by construction span the eigenstates below a certain energy
which has been set to 4~eV above the Fermi level in the present
calculation. The transformation results in the following set of WFs:
For each Pt we obtain 5 $d$-orbitals centered at the atom and a single
$\sigma$-orbital located at an interstitial site. For each hydrogen we
find an $s$-orbital, $|i\rangle$ ($i=1,2$), which is slightly
elongated towards the contacting Pt atom. We proceed by transforming
the hydrogen $s$-orbitals into bonding and anti-bonding combinations
$|b\rangle=(|1\rangle+|2\rangle)/\sqrt{2}$ and
$|a\rangle=(|1\rangle-|2\rangle)/\sqrt{2}$. $|a\rangle$ and
$|b\rangle$ are the only states with significant weight on the
molecule and provide two conductance channels well separated in
energy. The on-site energies are $\langle b|H|b\rangle=-6.4$~eV and
$\langle a|H|a\rangle=0.1$~eV relative to the Fermi level of the
metal.  By cutting all coupling matrix elements involving the bonding,
respectively, the anti-bonding state we can directly test their
individual contributions to the total conductance when interference is
neglected. The result is shown in Fig.~\ref{fig2}. The narrow peak
just below $-7$~eV is completely gone when the bonding state is
removed but is not affected by the absence of the anti-bonding state.
The peak is thus clearly due to transmission through the bonding
channel which is in good agreement with the calculated on-site energy
of $|b\rangle$. In the energy regime -6 -- -4 eV both the bonding and
anti-bonding states contribute to the transmission. For energies above
-3 eV the removal of the bonding state has little effect on the
transmission which must therefore be ascribed to the anti-bonding
state. A small exception to this is the narrow peak at -1 eV which is
caused by hybridization of the bonding state with Pt
$d_{z^2}$-orbitals on the contacting atoms. Overall we can conclude
that the peak at -2 eV and the $1G_0$-plateau which determines the
conductance are due to transmission through the anti-bonding state.

\begin{figure}[!b]
\includegraphics[width=0.62\linewidth]{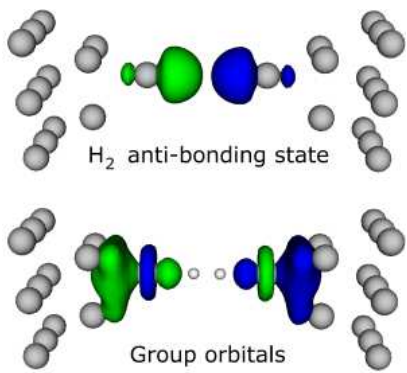}
\caption[cap.wavefct]{\label{fig3} Contour plots of the orbitals
  determining the transport properties of the hydrogen contact: the
  $\text{H}_2$ anti-bonding state, $|a\rangle$, and the corresponding
  left and right group orbitals, $|g_L\rangle,|g_R\rangle$. The left
  (right) group orbital has been constructed by applying the
  DFT Hamiltonian to $|a\rangle$ and then projecting onto the
  Wannier functions of the left (right) part of the contact.}
\end{figure}

The fact that the bonding state takes almost no part in the
transmission around the Fermi level allows us to describe the contact
by a resonant level model~\cite{newns69} with all parameters
determined from first principles. In the resonant level model we
consider a single level, $|a\rangle$, of energy $\varepsilon_a=\langle
a|H|a\rangle$ coupled to infinite leads via the matrix elements
$t_{k\alpha}=\langle k\alpha|H|a\rangle$, where $\{|k\alpha\rangle\}$
is a basis of lead $\alpha$. The model has served as starting point
for many more advanced studies such as shot noise, electron-electron
and electron-phonon interactions in resonant tunneling
systems~\cite{thielmann03,ng88,hyldgaard94}. A particularly useful
formulation of the model can be obtained if we introduce the group
orbital of lead $\alpha$ by
$|g_{\alpha}\rangle=c_{\alpha}P_{\alpha}H|a\rangle$, where
$P_{\alpha}$ is the orthogonal projection onto lead $\alpha$ and
$c_{\alpha}$ is a normalization constant. By supplementing the group
orbital by orthonormal states $\{|\tilde k\alpha\rangle \}$ we obtain
a new basis with the key property $\langle \tilde
k\alpha|H|a\rangle=0$ for all $\tilde k$. The level is thus coupled to
the lead via the group orbital only.  Since the contact is symmetric
$\langle g_L|H|a\rangle=\langle g_R|H|a\rangle\equiv V$.  The
imaginary part of the level self-energy, $\Delta_a$, is directly
related to the DOS of the group orbitals calculated with $V=0$:
$\Delta_a=\pi|V|^2(\rho^0_L+\rho^0_R)=(\Gamma_L+\Gamma_R)/4$.  The
real part of the self-energy is the Hilbert transform of $\Delta_a$.
For a symmetric contact we have $\rho^0_L=\rho^0_R\equiv\rho^0_g$ and
the transmission in Eq.~(\ref{eq.condformula}) takes the simple form
\begin{equation}\label{simpletrans}
T(\varepsilon)=2\pi^2|V|^2\rho^0_g(\varepsilon)\rho_a(\varepsilon).
\end{equation}
Since $\rho_a$ can be obtained from $\Sigma_a$ and $\varepsilon_a$
this expression shows that the transmission is determined by the three
quantities $\rho^0_g$, $V$ and $\varepsilon_a$.

\begin{figure}[!b]
\includegraphics[width=0.8\linewidth,angle=270]{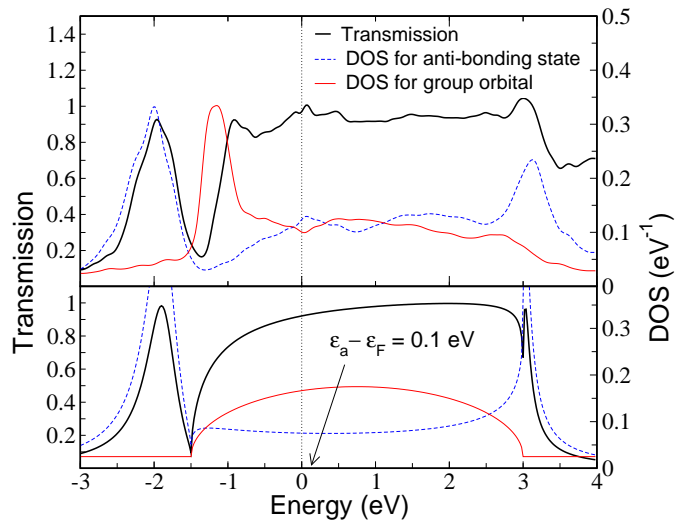}
\caption[cap.wavefct]{\label{fig4} Upper panel shows the transmission
  together with the projected density of states (DOS) for the
  $\text{H}_2$ anti-bonding state and the corresponding left group
  orbital. 
  The lower panel shows the same quantities obtained from the
  single-level model when 
  $\rho_g^0$ is approximated by a semi-elliptical band and we use the
  coupling, $V$, and on-site energy, $\varepsilon_a$, from the first
  principles calculation.}
\end{figure}

By applying the DFT Hamiltonian to $|a\rangle$ in the WF basis we
construct the group orbitals of the $\text{H}_2$ anti-bonding state.
Contour plots of the orbitals are shown in Fig.~\ref{fig3}. The group
orbital is mainly composed of the $d_{z^2}$-orbital of the apex Pt
atom and three $\sigma$-orbitals centered within the Pt pyramid.  We
calculate $\rho^0_g$ for the uncoupled system by cutting all coupling
matrix elements to $|a\rangle$.  The result is shown in the upper
panel of Fig.~\ref{fig4} together with $\rho_a$ and the transmission
function. The pronounced peak at -1 eV is due to the
$d_{z^2}$-orbitals on the apex Pt atoms.

If we neglect the narrow peak at $-1$~eV, $\rho^0_g$ can be described
by a semi-elliptical band on top of a flat background, see lower panel
of Fig.~\ref{fig4}.  The coupling and level energy can be directly
read off the Hamiltonian matrix and we find $V=1.9$~eV and
$\varepsilon_a=0.1$~eV relative to the Fermi level. It should be
noticed that the coupling which is relevant for the adsorption of the
hydrogen molecule to the contact is $\sqrt{2}V=2.7$~eV since the level
is coupled by $V$ to \emph{both} leads.  From these parameters we can
determine the Green's function for the level which in turn yields
$\rho_a$ and $T$. The result is summarized in the lower panel of
Fig.~\ref{fig4}. Based on the good agreement with the first principles
results we conclude that the simple model indeed gives a realistic
description of the system. It is then clear that the peak at $-2$~eV
represents the bonding combination between $|a\rangle$ and the Pt band
and that the $1G_0$-plateau forms because: (i) $\varepsilon_a$ lies
close to the Fermi level and well inside the relevant Pt band as
defined by the group orbital. (ii) the width of the renormalized level
($\Delta_a$) is comparable to the band width, i.e.  the limit of
strong chemisorption~\cite{newns69}.

The crucial point in the proposed mechanism is the strong
hybridization of the $\text{H}_2$ anti-bonding state with the Pt bands
around the Fermi level. This picture agrees well with the conventional
understanding of hydrogen dissociation on simple and transition metal
surfaces which has been established on the basis of DFT
calculations~\cite{hammer95,norskov_houmoller81}. The
bonding and antibonding states of a hydrogen molecule at a simple
metal surface are broadened and furthermore shifted down due to the
hybridization with the metal $s$- and $p$-states. During the
dissociation process the antibonding resonance crosses the Fermi level
and becomes gradually filled with the result that the
hydrogen-hydrogen bond is weakened. For the transition metal the same
general picture applies, but the hybridization with the $d$-states
further affects the antibonding resonance. The fact that the
antibonding state in the calculations for the bridging hydrogen
molecule between Pt contacts is close to the Fermi level is thus an
indication that the hydrogen-hydrogen bond is weakened by the coupling
to the metal in agreement with the resulting increase of the
hydrogen-hydrogen bond length. The values we find for the positions of
the bonding and antibonding molecular levels, $\varepsilon_b= -6.4$~eV, $\varepsilon_a=
0.1$~eV are in fact quite close to the ones used by Hammer and
N{\o}rskov~\cite{hammer95} ($\varepsilon_b= -7$~eV, $\varepsilon_a=1$~eV) to describe hydrogen
in the dissociative transition state on metals. This is in clear
contrast to the studies by Cuevas \emph{et
al.}~\cite{cuevas_heurich03} and Garc{\'i}a~\emph{et
al.}~\cite{garcia_palacios04} who consider the hydrogen molecule in
the bridging position to have almost the same bond legnth as the free
molecule and who report very large bonding-antibonding splittings of
23-24 eV which even exceed the DFT-PW91 value of 10.4 eV for a free
molecule. 

In summary, we have presented first principles conductance
calculations showing that a hydrogen molecule suspended between Pt
contacts can have a conductance close to $1G_0$.  Through a detailed
Wannier function analysis we have identified the conduction mechanism
as being due to a strong hybridization between the $\text{H}_2$
anti-bonding state and certain Pt bands. A resonant level model with all
parameters determined from the self-consistent DFT Hamiltonian was
shown to account for the important features of the first principles
transmission function.

We would like to thank J. van Ruitenbeek and D. Djukic for many inspiring
discussions. We acknowledge support
from the Danish Center for Scientific Computing through Grant No.
HDW-1101-05.


\bibliographystyle{apsrev}

\begin{thebibliography}{24}
\bibitem{calc} The surface is modeled as a slab with a thickness of 4
  atomic layers and the supercell contains $3\times 3$ atoms in the
  surface plane.  Both the Pt pyramids and the H atoms have been
  relaxed using a plane wave DFT code~\protect\cite{dacapo} to obtain
  the most stable contact geometry.  We use an energy cut-off of 25~Ry
  for the plane wave expansion and describe the ion cores by ultrasoft
  pseudopotentials~\protect\cite{vanderbilt90}. To treat exchange and
  correlation we use the PW91 functional~\protect\cite{pw91}. The
  Brillouin zone is sampled by a single $\bold k$-point along the
  contact axis ($z$-axis) and $4\times 4$ $\bold k$-points in the
  transverse plane.
\end{thebibliography}

\end{document}